\documentclass[aps, 
    a4paper,nofootinbib,twocolumn,%
]{revtex4} 
\usepackage[colorlinks,linkcolor=blue,citecolor=blue,urlcolor=blue ]{hyperref}
\usepackage{acronym}

\usepackage{multirow}
\usepackage{graphicx}
\graphicspath{{figs/}}
\usepackage{amsmath,amssymb}
\usepackage{braket}

\newcommand{\TRC}{MOE Key Laboratory of TianQin Mission, %
    TianQin Research Center for Gravitational Physics $\&$ School of Physics and Astronomy, %
    Frontiers Science Center for TianQin, %
    Gravitational Wave Research Center of CNSA, %
    Sun Yat-sen University (Zhuhai Campus), %
Zhuhai 519082, China}

\allowdisplaybreaks

\begin{document}

\title{Parameter estimation of stellar mass binary black holes in the network of TianQin and LISA}

\author{Xiangyu Lyu}
\author{En-Kun Li}
\thanks{Corresponding author: \href{mailto:lienk@mail.sysu.edu.cn}{lienk@mail.sysu.edu.cn}}
\author{Yi-Ming Hu}
\thanks{Corresponding author: \href{mailto:huyiming@mail.sysu.edu.cn}{huyiming@mail.sysu.edu.cn}}

\affiliation{\TRC}

\begin{abstract}
    We present a Bayesian parameter estimation progress to infer the stellar mass binary black hole properties by TianQin, LISA, and TianQin+LISA. 
    Two typical stellar mass black hole binary systems, GW150914 and GW190521 are chosen as the fiducial sources.
    In this work, we establish the ability of TianQin to infer the parameters of those systems and first apply the full frequency response in TianQin's data analysis.
    We obtain the parameter estimation results and explain the correlation between them.
    We also find the TianQin+LISA could marginally increase the parameter estimation precision and narrow the $1\sigma$ area compared with TianQin and LISA individual observations.
    We finally demonstrate the importance of considering the effect of spin when the binaries have a nonzero component spin and great deviation will appear especially on mass, coalescence time and sky location.
\end{abstract}



\maketitle
\acrodef{gw}[GW]{gravitational wave}
\acrodef{sbh}[sBH]{stellar-mass Black Hole}
\acrodef{bh}[BH]{black hole}
\acrodef{bbh}[sBBH]{stellar mass binary black hole}
\acrodef{mbhbs}[MBHBs]{massive black hole binaries}
\acrodef{mbhb}[MBHB]{massive black hole binaries}
\acrodef{emris}[EMRIs]{extreme-mass-ratio inspirals}
\acrodef{emri}[EMRI]{extreme-mass-ratio inspiral}
\acrodef{gbs}[GBs]{Galactic double white dwarf binaries}
\acrodef{gb}[GB]{Galactic double white dwarf binary}
\acrodef{sgwb}[SGWB]{stochastic gravitational-wave background}
\acrodef{psd}[PSD]{power spectral density}
\acrodef{snr}[SNR]{signal to noise ratio}
\acrodef{pe}[PE]{parameter estimation}
\acrodef{gr}[GR]{general relativity}
\acrodef{fim}[FIM]{Fisher information matrix}
\acrodef{tdi}[TDI]{time delay interferometry}
\acrodef{mhz}[mHz]{milli-Hertz}
\acrodef{mcmc}[MCMC]{Markov chain Monte Carlo}

\section{Introduction}

In the 2030s, multiple spaceborne \ac{gw} detectors, TianQin \cite{luo_tianqin:_2016}, LISA \cite{LISA:2017pwj} and Taiji \cite{Hu:2017mde}  will be launched and operational. They could detect \ac{gw} signals from various sources, including \ac{mbhbs}\cite{Wang:2019ryf}, \ac{emris}\cite{Fan:2020zhy}, \ac{gbs}\cite{Huang:2020rjf}, \ac{sgwb}\cite{Liang:2021bde}, and \ac{bbh} inspirals\cite{liu_science_2020}. As potential detectable sources for spaceborne detectors, \acp{bbh} have been detected by ground-based detectors (LIGO-Virgo-KAGRA) \cite{LIGOScientific:2016aoc}, with nearly 100 events reported  \cite{LIGOScientific:2018mvr, LIGOScientific:2021usb, LIGOScientific:2021djp}. Moreover, a fraction of these events can be detected by both TianQin \cite{liu_science_2020} and LISA \cite{sesana_prospects_2016}, which allows for joint observation (TianQin + LISA) and being possible multiband sources.



Accurate parameter estimation (PE) is crucial for characterizing the physics of GW sources. Parameters such as spin and mass can provide hints about the population properties of different formation channels. Many numerical simulations have shown that binaries formed in galactic fields should have spins preferentially aligned with the angular momentum 
\cite{1993MNRAS.260..675T,Wong:2013vya,Belczynski:2017gds,Zaldarriaga:2017qkw,Stevenson:2017tfq,Gerosa:2018wbw},
while the spin tends to have a random orientation when binaries are formed dynamically
\cite{PortegiesZwart:2002iks,Rodriguez:2013mla,Antonini:2016gqe,Robson:2018ifk,Gerosa:2021mno}. 
The spin of a binary also correlates with its mass, with more unequal-mass systems having a more significant effective spin and the GWTC-3 dataset supports the correlation between spin and mass \cite{Franciolini:2022iaa}. 
The value of the mass also plays an important role in distinguishing the \ac{bbh} system formation channel, the event GW190521 \cite{LIGOScientific:2020iuh} indicates \ac{bbh} could lie in the mass gap~\cite{LIGOScientific:2020ufj,Romero-Shaw:2019itr,Gayathri:2020coq,Barrera:2022yfj}, which is the mass gap of stellar-mass black holes (sBHs) predicted by pair-instability supernovae (PISNs). 
Notice that different teams report different masses for GW190521~\cite{nitz_gw190521_2021,fishbach_minding_2020}, adding extra layers of complexity to its formation channel. Therefore, achieving precise PE results is vital for increasing scientific understanding and improving our knowledge of \acp{bbh}~\cite{Sesana:2017vsj,Seto:2016wom}.


Since the \ac{gw} detection of the first \ac{bbh} systems, many works have demonstrated how spaceborne \ac{gw} detectors can improve precision on the physical parameters, and how a joint observation TianQin and LISA can further improve the science.
\citet{sesana_prospects_2016} demonstrated that LISA could provide strict constraints on the \ac{bbh} mass with the relative error to be $\sim 10^{-6}$. Besides spin, parameters such as sky location and coalescence time can be estimated with high precision, less than $1 \, \rm deg^2$ and less than $10$ seconds respectively. This level of precision in time and space enables the electromagnetic (EM) and ground-based detectors' follow-up observation. Furthermore, 
TianQin has also shown similar capabilities in PE, with joint observation allowing for significant improvements in coalescence time, sky localization, and chirp mass
~\cite{liu_science_2020,Liu:2021yoy}. As an ideal type of multiband source, a precise estimation of the sky location of \ac{bbh} system could also enable us to constrain the Hubble constant
\cite{Kyutoku:2016zxn, DelPozzo:2017kme}, with even better precision achievable through joint observations~\cite{Zhu:2021bpp}. 

Previous studies also indicate that precise measurement on \acp{bh} with spaceborne \ac{gw} missions is crucial for analyzing the environment and the astrophysical formation mechanism of \acp{bbh}~\cite{Benacquista:2011kv,Antonini:2012ad,Postnov:2014tza,Bird:2016dcv,Samsing:2017xmd,Gerosa:2019dbe,Su:2021dwz,Klein:2022rbf}, as well as the testing of the general relativity (GR)~\cite{Berti:2004bd,Vallisneri:2007ev,Vitale:2016rfr,Barausse:2016eii,Chamberlain:2017fjl,Gnocchi:2019jzp,Carson:2019rda,Toubiana:2020vtf,Nakano:2021bbw,LISA:2022kgy}. 
However, many of the aforementioned works rely on the method of the \ac{fim}, which is lightweight in terms of computation demands, with the cost of neglecting details~\cite{Vallisneri:2007ev,Vallisneri:2011ts}.
Recent works have started to utilize the Bayesian inference to perform the PE in \ac{bbh} inspirals~\cite{marsat_exploring_2021,Toubiana:2020cqv,Toubiana:2022vpp,buscicchio_bayesian_2021} and all of them do not consider the noise simulation. 
In our work, we choose to use Bayesian inference tools like the \ac{mcmc} methods to obtain the posterior distribution on the parameter space, in the hope the obtain a more accurate and reliable understanding of future data analysis ability of spaceborne \ac{gw} missions.

Although many mature pipelines have been developed in the context of ground-based \ac{gw} detectors \cite{Ashton:2018jfp,Veitch:2014wba,Lange:2018pyp}, spaceborne missions face different challenges.
For example, due to the long-lasting nature of the source and the motion of the spaceborne \ac{gw} detectors, the response function variation to \ac{gw} signals and rotation of the constellation cannot be ignored over the long lifetime of the binaries. Also, the commonly used response model that works in the low-frequency limit \cite{cutler_angular_1998} is not applicable when \ac{bbh} inspirals approach high-frequency regions.
To obtain a realistic and comprehensive understanding of precisely how future spaceborne missions can constrain \ac{bbh} parameters, we incorporate the \ac{tdi} response described in \cite{marsat_fourier-domain_2018, marsat_exploring_2021}. This response model is available in full frequency region and has been incorporated into the analysis for TianQin.

In our work, we aim to develop software to perform Bayesian inference on \ac{bbh} inspiral with simulated data for spaceborne \ac{gw} detectors, TianQin, while taking the full frequency \ac{tdi} response into consideration, both in generating data and parameter estimation. Our method could be easily applied to other \ac{gw} sources. Here, we apply this software to two of the most studied \ac{gw} events, GW150914\cite{LIGOScientific:2016aoc} and GW190521\cite{LIGOScientific:2020iuh}.
The first for being the first ever detection and also among the loudest event; the second for being among the heaviest systems.
For the Bayesian inference algorithm, we use the affine-invariant sampling software \texttt{emcee} \cite{foreman-mackey_emcee_2013} to obtain the posterior distribution.

The paper is organized as follows. In Sec.\ref{title:technology}, we give a brief description of the theoretical foundations for the \ac{bbh} systems, including the waveform model we adopt, the \ac{tdi} response, and the concepts of the TianQin mission.
In Sec.\ref{title:PE method}, we present the Bayesian framework as well as the implementation of the \ac{pe} methods. In Sec.\ref{title:results}, we provide our PE results for the two events and discuss the impact of spin mismodeling on those results.
In Sec.\ref{title:conclusion}, we summarize our findings and discuss possible future works.
Throughout the work, we use the geometrical units $(\rm G = c = 1)$ unless otherwise stated.


\section{Theoretical basics}\label{title:technology}

\subsection{sBBH system}\label{title:sBBH system}

The spaceborne \ac{gw} detectors could only observe the inspiral stage of these binaries, and no confident observation of orbital eccentricity has been observed with ground-based facilities. 
For simplicity, we assume the binaries follow quasicircular orbits, therefore the contribution from other subdominant harmonics can be neglected and the 22 mode is sufficient to describe the waveform. 
In our work, we adopt the \texttt{IMRPhenomD} waveform \cite{Khan:2015jqa, Husa:2015iqa}. 
The choice is made so that our method is general enough that can be easily applied to other types of waveforms.  
Notice that \texttt{IMRPhenomD} assumes aligned spin so it uses only two parameters to describe the spin parameters. 
For the adoption of more general waveforms, one needs to consider the spin-precession effect, which will increase the complexity of the waveform \cite{marsat_exploring_2021}.
In this frame, a \ac{bbh} system can be characterized by four intrinsic parameters: two component masses $(m_1, m_2)$ and two component dimensionless spins $(\chi_1,\chi_2)$; and seven extrinsic parameters: luminosity distance $D_L$ of the source, inclination angle $\iota$ that describes the angle between the orbital angular momentum with respect to the line of sight, polarization angle $\psi$, 
coalescence time and phase $(t_c, \phi_c)$ and the ecliptic longitude and ecliptic latitude $(\lambda,\beta)$ in the solar-system barycenter (SSB). We set a group of parameters that are used for estimation and they are $\boldsymbol{\theta} = (D_L,M_c,\chi_a,\chi_l,t_c,\eta,\iota,\lambda,\beta,\psi,\phi_c)$, where $M_c = \frac{(m_1m_2)^{3/5}}{(m_1+m_2)^{1/5}}$ is the chirp mass, $\eta = m_1m_2/(m_1+m_2)^2$. And due to $\chi_1$ and $\chi_2$ being highly correlated, we reparametrize them into $\chi_a,\chi_l = (\chi_1+\chi_2)/2, (\chi_1-\chi_2)/2$ to remove some correlation and use them to perform the parameter estimation. In the \texttt{IMRPhenomD} waveform model, \ac{bbh} can be described by $\tilde{h}(f) = \tilde{h}_+(f) - {\rm i}\tilde{h}_{\times}(f)$, where 
\begin{equation}
    \begin{aligned}
    \tilde{h}_+(f) =& \frac{1 + \cos^2\iota}{2}
    \frac{M_c^{5/6}}{\pi^{2/3}D_L}f^{-7/6} \exp({\rm i}\Psi(f)),
    \\
    \tilde{h}_{\times}(f) =& -{\rm i} \cos{\iota} 
    \frac{M_c^{5/6}}{\pi^{2/3}D_L}f^{-7/6}\exp({\rm i}\Psi(f)).
    \end{aligned}
    \label{equ:imr waveform}
\end{equation}
More details about the phase $\Psi(f)$ can be seen in \citet{Khan:2015jqa}.


Some parameters may have similar effects on the waveform, therefore leading to degeneracies between parameters. 
For example, in our work, we focus on the inspiral stage and near-circular systems and use the waveform only in 22 mode. 
Therefore, the inclination angle $\cos{\iota}$ and luminosity distance $D_L$ are strongly correlated as they both appear in the amplitude with combinations like $(1+\cos^2{\iota})/2D_L$ and $\cos{\iota}/D_L$ in Eq.(\ref{equ:imr waveform}). The degeneracy can be broken if the polarization can be measured accurately since their combinations differ in different polarizations \cite{marsat_exploring_2021}, or if higher modes can be observed as $D_L$ and $\iota$ affect the amplitudes through different combinations \cite{Yang:2022tig}.
Another example is the two spins $(\chi_a,\chi_l)$. 
The  leading orders of $\Psi(f)$  that contain spin parameters are the 1.5 and 2 order post-Newtonian (PN) terms, which are $ \frac{113\delta}{3}\chi_l + \left( \frac{113}{3} - \frac{76\eta}{3}  \right) \chi_a, 
\rm and \left( -\frac{405}{8} + 200\eta \right) \chi_l^2 - \frac{405}{4}\delta\chi_a\chi_l + \left( -\frac{405}{8} + \frac{5\eta}{2} \right)\chi_a^2$ respectively, where $\delta=(m_1-m_2)/(m_1+m_2)$.

\subsection{Detector response}
\label{title:response_signal}

For some science cases studies for LISA\cite{sesana_prospects_2016,Sesana:2017vsj} and TianQin\cite{liu_science_2020,Liu:2021yoy}, had not adopted \ac{tdi}, since it does not significantly alter \ac{snr} while greatly complicates the calculations. However, in reality, since the laser phase noise can be orders of magnitude higher than the \ac{gw} signals, one has to rely on the TDI combination of channels to cancel out the laser phase noise \cite{armstrong_time-delay_1999, dhurandhar_algebraic_2002, estabrook_time-delay_2000, tinto_cancellation_1999}. Various TDI schemes can be constructed with different combinations of measurement readouts from satellites.
In particular,  we adopt the A, E, T \cite{prince_lisa_2002} combination, with A and E reflecting the two \ac{gw} polarizations and the T channel is a noise-monitoring channel. 

In this work, we utilized the analytical full frequency response described in \cite{marsat_fourier-domain_2018, marsat_exploring_2021} for the purpose of ensuring speed. The TianQin TDI response can be constructed by combining the delayed signal-link observables. The signal-link observables denoted as $y_{slr}$, where $y_{slr} = (\nu_r - \nu_s) / \nu $ represents the \ac{gw}-induced laser frequency shift between the transmitting satellite $s$ and the receiving satellite $r$ along the link $l$. 
The equation below defines the observable $y_{sr}$:
\begin{equation}
  y_{sr} = \frac{1}{2} \frac{\hat{n}_l \otimes \hat{n}_l}{1-\hat{k}\cdot \hat{n}_l}:\left[h\left(t-L-\hat{k}\cdot\vec{p}_s\right)-h\left(t-\hat{k}\cdot \vec{p}_r\right)\right], 
    \label{yslr}
\end{equation}
where $h$ denotes the transverse-traceless metric perturbation \cite{Vallisneri:2004bn,Krolak:2004xp}, $L$ is the armlength, $\hat{n}_l$ is the unit vector from satellite $r$ to $s$, the vector $\hat{k}$ is the unit vector of \ac{gw} propagation, and $\{\vec{p}_r,\, \vec{p}_s\}$ represent the gravitational wave propagation direction and the positions of two satellite, respectively. Here we assume a rigid armlength between satellites and the delay is the same in each link. The Michelson $X$ channel \cite{vallisneri_synthetic_2005} is composed of: 
\begin{equation}
    \begin{aligned}
    X =& y_{31}+y_{13,L} + (y_{21} + y_{12,L})_{,2L} 
    \\
    & - (y_{21} + y_{12,L}) - (y_{31} + y_{13,L})_{,2L},
    \end{aligned}
    \label{equ:x channel}
\end{equation}
where $y_{sr,nL} = y_{sr}(t - nL)$ and the other Michelson observables $Y$, $Z$ are obtained through cyclic permutation. 
The TDI channels $A$, $E$, and $T$ are orthonormalized. The $A$ and $E$ channels contain the \ac{gw} signal, while the $T$ channel is insensitive to \ac{gw} signals.
\begin{equation}
    \begin{aligned}
        &&A &= \dfrac{1}{\sqrt{2}}(Z-X), &\\
              &&E &= \dfrac{1}{\sqrt{6}}(X-2Y+Z),&\\
                    &&T &= \dfrac{1}{\sqrt{3}}(X+Y+Z).&
        \label{AET}
    \end{aligned}
\end{equation}
In the TDI channels, we denote the signal-link observable $\Tilde{h}_{slr}$ to be the Fourier transform of $y_{slr}$, and it can be explained by the \ac{gw} $\tilde{h}(f)$ in frequency domain and transfer function $\mathcal{T}_{slr}$:
\begin{equation}
    \tilde{y}_{slr} = \mathcal{T}_{slr}\tilde{h}(f)
\end{equation}
Furthermore, in the leading order of timescale separation~\cite{marsat_fourier-domain_2018}, the transfer function $\mathcal{T}_{slr}$ could be expanded to:
\begin{equation}
    \mathcal{T}_{slr}(f) = G_{slr}(f,t(f)),
    \label{equ:T22 1}
\end{equation}
where 
\begin{gather}
    \begin{aligned}
        G_{slr}(f,t) =&
        \dfrac{{\rm i}\pi fL}{2} \mathrm{sinc}[\pi fL(1-k\cdot n_l)] 
        \\
        & \cdot \exp[{\rm i}\pi f(L+k\cdot(p_r+p_s))]\ n_l\cdot P\cdot n_l,
    \end{aligned}     
    \label{equ:G22 and tf}
\end{gather}
where $P$ is the polarization tensor \cite{marsat_exploring_2021}. 
As the \ac{bbh} inspiral waveform consists of slowly varying phase and amplitude, which can be written as $\tilde{h}(f) = A(f)e^{-{\rm i}\Psi(f)}$. In Eq.~\ref{equ:T22 1}, the time-frequency correspondence can be obtained by waveform phase $\Psi(f)$, and $t(f) = - \frac{1}{2\pi} \frac{{\rm d} \Psi(f)}{{\rm d}f }$.

\subsection{TianQin mission}
\label{title: TQ mission}


TianQin \cite{TianQin:2015yph} is a \ac{gw} observatory designed for spaceborne operation, consisting of a constellation of three Earth-orbiting satellites arranged in an approximately equilateral triangular formation with an armlength of about $\sqrt{3}\times 10^5$km. The norm of the orbital plane is oriented towards the double white dwarf (DWD) system J0806 \cite{beuermann_identification_1999, Israel:2002gq}. Due to the specific orbit configuration, TianQin operates in a three month on, three month off scheme, so that the thermal load imposed by direct sunlight entering the telescopes can be mitigated.



TianQin is anticipated to observe various types of \ac{gw} sources, and many works have utilized the TianQin to explore physics and cosmology. For example, GBs are expected to be the most abundant detectable source for TianQin, and the detection of more GB events can help test formation models and analyze the mass distribution of our Galaxy~\cite{Huang:2020rjf}. Additionally, research indicates that TianQin could observe the~\ac{mbhbs} \cite{Wang:2019ryf} and constrain the source parameters. Moreover, TianQin is expected to observe the \ac{emris} \cite{Fan:2020zhy}, providing opportunities to test general relativity through their \ac{gw} characteristics. The researches \cite{liu_science_2020,Liu:2021yoy} based on TianQin, have demonstrated its ability to detect \ac{bbh} system and precisely estimate the source parameters. The \ac{sgwb} consists of a multitude of GW signals, the research \cite{Liang:2021bde} has shown that TianQin can place constraints on the source parameters of the \ac{sgwb}.
In terms of cosmology, by utilizing the sky location information, TianQin can employ \ac{bbh},\ac{mbhbs}, and \ac{emris} to constrain the Hubble constant \cite{Zhu:2021aat, Zhu:2021bpp} and contribute to our understanding of the expanding universe. 



\begin{figure*}[htbp]
    \includegraphics[width=0.98\linewidth]{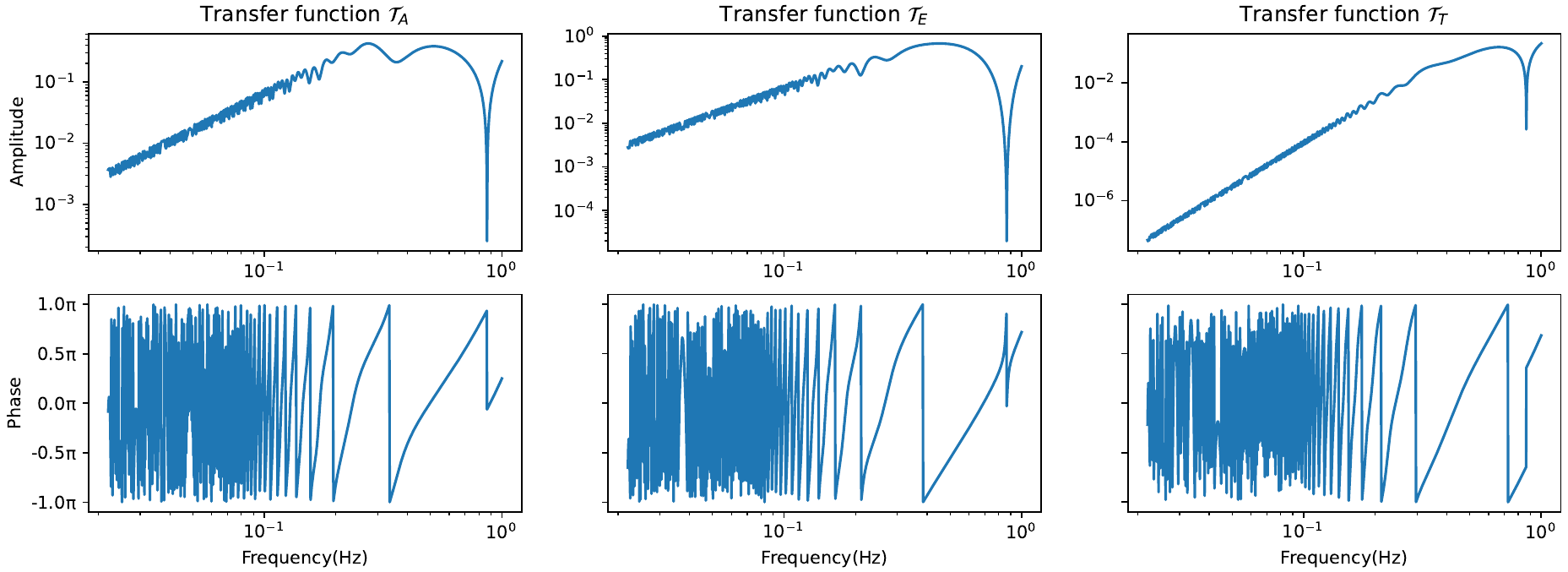}
    \caption{The frequency domain transfer function under TianQin TDI channels. }
    \label{fig:transfer AET}
\end{figure*}

As mentioned in Sec.\ref{title:response_signal}, sharing with a same constellation, the spaceborne response signals in each channel can be obtained as $\Tilde{h}_i = \mathcal{T}_i\Tilde{h}(f)$, where $\mathcal{T}_i(f)$ is the transfer function for each channel ($i = A, E, T$). 
 In Fig.~ \ref{fig:transfer AET}, we demonstrate the transfer functions of TianQin's three TDI channels. 
For both TianQin and LISA missions, their technical requirements lead to estimated accuracies as follows: For TianQin \cite{TianQin:2015yph}, the position noise $\sqrt{S_x} \approx 1 \rm pm/Hz^{1/2} $ and the residual acceleration measurements have a value of $\sqrt{S_a} \approx 10^{-15} \rm ms^{-2}/Hz^{1/2}$. The \ac{psd} for LISA can also be described with the similar scheme \cite{LISA:2017pwj}, and the noise parameters in PSD for LISA can be expressed as follows: the acceleration noise is $\sqrt{S_{\rm Acc}} \approx \rm 3\times10^{-15}m^{-2}/Hz^{1/2}$ and the interferometric measurement is $\sqrt{S_{\rm IMS}} \approx \rm 12 pm/Hz^{1/2}$.
The formulas for the noise PSD in the TDI channels ($A, E, T$) are as follows:
\begin{equation}
    \begin{aligned}
        S_n^A = S_n^E = & \, 8 \sin^2{\omega L} \bigg[ 4(1+\cos{\omega L}+\cos^2{\omega L})S_{\rm acc} \\
        & + (2+\cos{\omega L})S_{\rm pos} \bigg], \\
        S_n^T = & \, 32\sin^2{\omega L}\sin^2{\frac{\omega L}{2}}\left[4\sin^2{\frac{\omega L}{2}}S_{\rm acc}+S_{\rm pos}\right],
    \end{aligned}
    \label{AET_PSD}
\end{equation}
where the unfolding of noise parameters $S_{\rm acc}$ and $S_{\rm pos}$ in the case of TianQin and LISA can be seen in Refs. \cite{luo_tianqin:_2016} and \cite{noauthor_lisa_nodate}.

\section{Parameter estimation methodology}\label{title:PE method}

In this study, our primary focus is on the \ac{bbh} parameter estimation rather than detection. The successful detection method used by ground-based detectors is matched filtering, which requires a set of waveform templates~\cite{Martel:1999zx}. However, this method poses significant challenges for space-based gravitational wave detectors, and \citet{Moore:2019pke} has demonstrated that for LISA, it requires a template bank on the order of $10^{30}$ in size for matched filtering. The size of the template bank far exceeds acceptable computational limitations. There have been efforts to address this issue, like using the semicoherent search methods \cite{Bandopadhyay:2023gkb} or adopting the archive search methods~\cite{Wang:2023dgm,Klein:2022rbf,Gerosa:2019dbe,sesana_prospects_2016,Ewing:2020brd}, which utilize information from ground-based observations to reduce the size of the template bank. 

\subsection{Bayesian inference theory}
\label{bayesian inference}

We work on the Bayesian framework for the PE, and according to the Bayes theorem, the posterior probability distribution $p(\boldsymbol{\theta}|D)$ of the source parameters $\boldsymbol{\theta}$ can be expressed as:
\begin{equation}
    p(\boldsymbol{\theta}|D) = \dfrac{p(D|\boldsymbol{\theta})p(\boldsymbol{\theta})}{p(D)},
    \label{equ:bayes equation}
\end{equation}
where $p(\boldsymbol{\theta})$ represents the prior information obtained prior to the detection,  $p(D|\boldsymbol{\theta})$ is the likelihood, and $p(D)$ denotes the evidence or marginalized likelihood. And our work only concerns parameter estimation and does not perform model selection, we can safely treat $p(D)$ as a normalization constant for the posterior $p(\boldsymbol{\theta})$. Therefore, the posterior probability of the source parameters is proportional to the product of the prior and the likelihood $\mathcal{L}=p(D|\boldsymbol{\theta})$:
\begin{equation}
    p(\boldsymbol{\theta}|D) \propto \mathcal{L} \times p(\boldsymbol{\theta})
    \label{equ:posterior}
\end{equation}
Here we introduce the inner product,
\begin{equation}
    (D|h) = 4\Re\int_0^{+\infty}{\rm d}f\frac{\Tilde{D}^*(f)\Tilde{h}(f)}{S_n(f)},
        \label{equ:inner product}
\end{equation}
where $S_n(f)$ represents the one-sided PSD of detector noise, $\Tilde{D}^*(f)$ is the complex conjugate of the observation data in the frequency domain. The Bayesian inference method was employed to extract the physical information from the data. This method involves comparing the GW templates $\Tilde{h}^{'}(f)$ with the data $D$. For stationary Gaussian noise, the logarithm of the likelihood function $\mathcal{L}$ of the \ac{gw} signal takes the following form.
\begin{equation}
    \begin{aligned}
        \ln{\mathcal{L}} &\propto -\frac{1}{2}(D - h(\boldsymbol{\theta})|D - h(\boldsymbol{\theta})), \\
        &= -\frac{1}{2}\left[ (D|D) + (h(\boldsymbol{\theta})|h(\boldsymbol{\theta})) -2(h(\boldsymbol{\theta})|D)  \right]
        \label{equ:likelihood ratio} \\
    \end{aligned}
\end{equation}
The term $(D|D)$ is independent of the parameters $\boldsymbol{\theta}$ and can be regarded as a constant in the likelihood calculation.

For a \ac{gw} detector, the optimal signal-to-noise ratio (SNR) $\rho$ for a GW signal $h(t)$ is defined as the square root of the inner product $(\cdot|\cdot)$,
\begin{equation}
    \rho^2 = \sqrt{(h|h)}= \sqrt{4\Re\int_0^{+\infty}{\rm d}f\frac{\Tilde{h}^*(f)\Tilde{h}(f)}{S_n(f)}}.
    \label{equ:inner product}
\end{equation}
If the signal was detected by multiple detectors, the total SNR could be derived by the root sum square of the SNR from individual detectors. For example, in the case of joint observations, the total squared SNR can be obtained by the sum of the individual detector (e.g., TianQin + LISA),
\begin{equation}
    \rho=\sqrt{\rm \rho_{TianQin}^2+\rho_{LISA}^2}
    \label{equ:total SNR}
\end{equation}

\subsection{Implementation}\label{Implementation}

Since there is no actual observation data of spaceborne GW interferometry available, we simulate \ac{bbh} inspiral data for our analysis. As described in Sec.~\ref{title:response_signal}, the TianQin and LISA observation data needed to be constructed from the $A$, $E$, and $T$ channels. In order to make a fair demonstration of the ability to constrain the source parameters for TianQin and LISA and also consider a reasonable computing time of PE, we choose the observation period for TianQin and LISA are both two years.
During the observation period, we assume that the TianQin satellite shut down during the 3rd$\sim$6th, 9th$\sim$12th, 15th$\sim$18th, 21st$\sim$24th months. As a result, the data $D(f)$ can be divided into eight parts corresponding to the remaining time periods: $D_{0-3}(f), D_{6-9}(f), D_{12-15}(f), D_{18-21}(f)$. Additionally, using the time-frequency mapping function \cite{cutler_gravitational_1994}, we can determine the starting and ending frequencies for each data segment.
Therefore, the 3+3 observation mode results in multiple three-month data segments.
Under this situation, we denote the TianQin data as $D_{\rm TianQin}: [D_{0-3},D_{6-9},D_{12-15},D_{18-21}]$.
The likelihood of these data segments for the source parameter ${\boldsymbol{\theta}}$ can be expressed as,
\begin{equation}
    \begin{aligned}
        \ln\mathcal{L}(D_{\rm TianQin}|\boldsymbol{\theta}) =& \ln\mathcal{L}(D_{0-3}|\boldsymbol{\theta}) + \ln\mathcal{L}(D_{6-9}|\boldsymbol{\theta}) 
        \\
        & +\ln\mathcal{L}(D_{12-15}|\boldsymbol{\theta}) +\ln\mathcal{L}(D_{18-21}|\boldsymbol{\theta}),
        \label{equ:TQ lnlikelihood}
    \end{aligned}
\end{equation}


For a network of multiple detectors, such as the TianQin+LISA network, we can express the joint likelihood as the product of individual likelihoods. Consequently, the logarithm of the joint likelihood becomes
\begin{equation}
    \begin{aligned}
        \ln\mathcal{L}_{\rm joint} =& \ln\mathcal{L}_{\rm TianQin} + \ln\mathcal{L}_{\rm LISA} \\
        =& -\dfrac{1}{2} \bigg[ (D_{\rm TianQin} - \Tilde{h}^{'}(\boldsymbol{\theta})|D_{\rm TianQin} - \Tilde{h}^{'}(\boldsymbol{\theta})) 
        \\
        & + (D_{\rm LISA} - \Tilde{h}^{'}(\boldsymbol{\theta})|D_{\rm LISA} - \Tilde{h}^{'}(\boldsymbol{\theta})) \bigg] 
        \label{equ:joint_lnlikelihood}
    \end{aligned}
\end{equation} 


Bayesian inference theorem provides a means to estimate the posterior distribution of parameters using their prior distribution $p(\boldsymbol{\theta})$ and likelihood $\mathcal{L}$. However, the dimension of the waveform parameters $\boldsymbol{\theta}$ could up to be $11$ in our study, and exploring high-dimensional posterior distribution will be computationally intensive. To address this, Markov chain Monte Carlo\cite{metropolis_equation_1953,hastings_monte_1970} methods have been widely adopted as efficient sampling algorithms to approximate the posterior distribution by counting the probability density of the sample points. In this study, we employ a program implementation of the affine invariant ensemble sampler for MCMC \texttt{emcee} \cite{foreman-mackey_emcee_2013}. In this software, several sampling algorithms are provided. Besides the default move algorithm 'stretch move', we consider using the other different move algorithm when generating the sample points. As we find they could help get more independent points with less computational consumption.

\section{Results}\label{title:results}

\begin{table*}[htbp]
    \centering
    \caption{ Parameters of the GW150914-like and GW190521-like case.}
    
    \label{tab:pars_table}
    \begin{tabular}{lccc} 
        \hline
        \textbf{Parameters} & \textbf{Symbols} & \textbf{~GW150914-like~} & \textbf{~GW190521-like}\\
        \hline
        Chirp mass $(M_{\odot})$ & $M_c$ & 32 & 88\\
        Symmetric mass ratio & $\eta$ & 0.24 & 0.23\\
        Primary spin, secondary spin & $(\chi_1, \chi_2)$ & (0.15, 0.05) & (0.69,0.73) \\
        \hline
        Inclination angle (rad) & $\iota$ & $3\pi/4$  &  $\pi/6$  \\
        Luminosity distance(Mpc) & $D_L$ & 100 & 900\\
        Coalescence time(sec)  & $t_c$ & $6.31\times 10^7$ & $6.31\times 10^7$  \\
        Ecliptic longitude  (rad) & $\lambda$ & 2.50 &5.66 \\
        Ecliptic latitude angle (rad) & $\beta$ & -1.17&-0.055\\
        Polarization angle (rad) & $\psi$ & $\pi/3$ & $2\pi/3$ \\
        Coalescence phase (rad) & $\phi_c$ & 0 & 0 \\
        \hline
    \end{tabular}
\end{table*}

In this work, we simulate the \ac{bbh} inspiral signals with parameter values from two detected events reported by the LIGO-Virgo Collaboration, namely GW150914 and GW190521, as fiducial systems.
We then perform parameter estimation on TianQin, LISA, and joint observation.
GW150914 is chosen as it is the first confirmed \ac{gw} event and therefore also the most thoroughly studied one.
GW190521 stands out as the component black hole's mass falls in the higher mass gap predicted by the PISN  \cite{Fowler:1964zz,Barkat:1967zz,Bond:1984sn,ober1983evolution,Heger:2002by,Woosley:2007qp,Chen:2014aba,10.1093/mnras/stv3002}, and the merger remnant mass $142_{-16}^{+28} \rm M_{\odot}$ made it the first confirmed intermediate-mass black hole. 

To perform parameter estimation, we first employ the {\texttt{IMRPhenomD}} waveform model to generate the \ac{gw} signal and compute the TDI response of the TianQin and LISA to obtain the response signals. Then, we set a flat prior distribution for all the parameters,
For the sampling process, we adopt the \texttt{emcee} realization of the affine invariant sampling \ac{mcmc} to efficiently sample in the parameter space. 

We first demonstrate the procedure we adopt to simulate the \ac{gw} signals in Sec.\ref{subtitle: data simulation}. 
The simulation signals are displayed and the corresponding sources' parameters are listed. The set of the prior range for all source parameters is shown in Sec.\ref{title:mock data prior}. 
The posterior distribution of two sources' parameters, GW150914-like and GW190521-like, is shown in Sec.\ref{title:PE results} in the first two parts with the form of a `corner plot' \cite{foreman-mackey_cornerpy_2016}. The PE results of TianQin, LISA, and TianQin+LISA are plotted together with different colors, and the difference in the posterior distribution can be seen from the contour plots. Almost all of the parameters we search over are significantly constrained compared to their respective priors. 
In the third part of Sec.\ref{title:PE results}, we discuss the effect of spin mismodeling when the \ac{gw} sources have nonzero spin. As the joint observation is an optimal situation and it actually performs better than the individual (TianQin/LISA). We utilize the zero spin waveform and the PE method based on joint observation to obtain the posterior distribution of all source parameters except spin. The PE results are plotted together, in order to clearly show the divergence of posterior distribution caused by zero-spin waveform and nonzero-spin waveform in the under the joint observation.

\subsection{GW data simulation}
\label{subtitle: data simulation}
For the simulation of data, we adopt the mean value presented in the GWTC-3 catalog \cite{LIGOScientific:2021djp} as injected parameters.
However, considering the detection ability \cite{Moore:2019pke} of the detectors, we make adjustments to the luminosity distance and sky location of the injected sources to ensure that the SNR of both sources meets the threshold of $12$ \cite{Moore:2019pke}, otherwise the \ac{snr} in TianQin and LISA might be too low to be detected. Here we assume that the spins of the binary components are either aligned or antialigned with the orbital angular momentum.  Consequently, we use two spin parameters $\chi_1$ and $\chi_2$ to represent the spin magnitude of the two components. All the source parameters are given in Table.~\ref{tab:pars_table} alone with the true values. The intrinsic parameters, mass and spin are listed in the top half of the table and the other extrinsic are displayed in the lower half for clarity. 


In our work, several assumptions were made in the analysis. Firstly, the noise realization $n$ was assumed to be zero in the simulated observation data for $D_{\rm TianQin}$ and $D_{\rm LISA}$. This assumption is commonly adopted in previous works \cite{marsat_exploring_2021,Toubiana:2020cqv}, and to make a meaningful comparison we also do not include noise in the simulated data.
Furthermore,  this study aims to assess the parameter estimation precision with \ac{mcmc}, and compare it with methods adopted in science case studies, like the \ac{fim}.
The inclusion of noise would cause the maximum \emph{a posteriori} to deviate from the injected value,  which will not appear in the \ac{fim} analysis.
Secondly, the systematic error caused by waveform mismodeling was neglected,
and we assume the waveforms accurately describe the GW signals. The selected waveform model, \texttt{IMRPhenomD}, was chosen based on the sensitive frequency band of TianQin and LISA, corresponding to the inspiral stage of \ac{bbh} systems. 
Thus, the joint likelihood can be approximated as the product of individual likelihoods, as described in Eq.~\eqref{equ:TQ lnlikelihood}. In the case of joint observation, the logarithmic likelihood is given by Eq.~\eqref{equ:inner product}.

\begin{figure}[htbp]
    \includegraphics[width=0.495\textwidth]{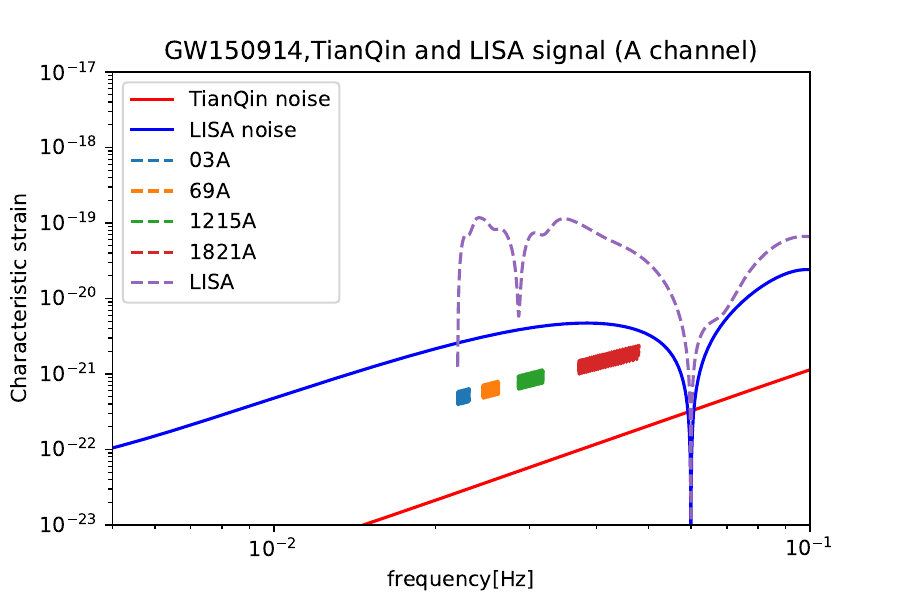}
    \includegraphics[width=0.495\textwidth]{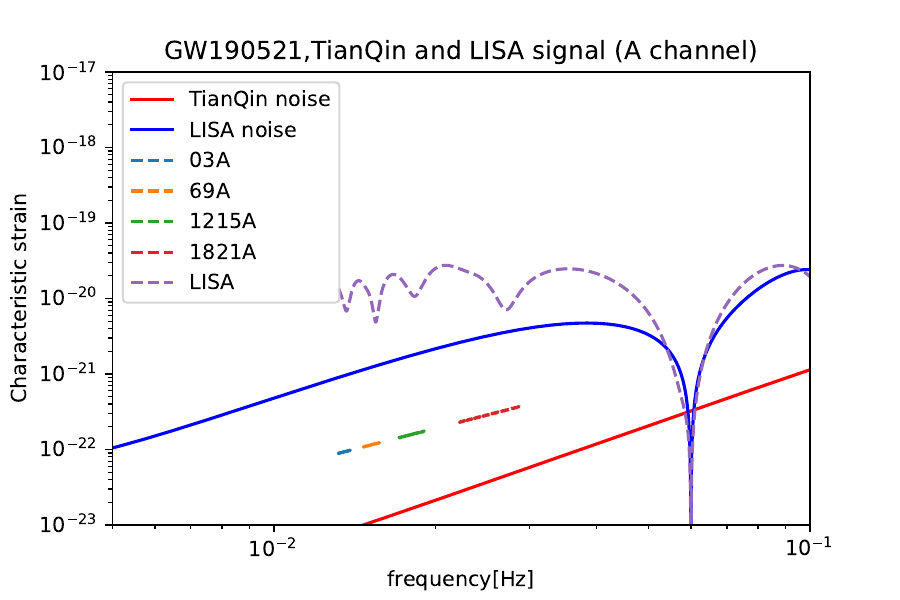}
    \caption{ The observation data and noise counterpart of GW150914 and GW190521 for TianQin and LISA are displayed in the upper and lower subplot. The purple dashed line and the other four parts dashed lines represent the characteristic strain $2f|\Tilde{h}(f)|$ observed by LISA and TianQin in 3+3 mode. The blue and red solid lines are the noise of the LISA and TianQin $\sqrt{f  S_n(f)}$.}
    \label{fig:mock_data} 
\end{figure}


TianQin and LISA's characteristic strains for GW150914 and GW190521 and noise in A/E channel are shown in Fig.~\ref{fig:mock_data}. Due to the observation scheme of ``three months on, three months off" for TianQin, the response signals are discontinuous. 
In the upper plot of Fig.~\ref{fig:mock_data}, as we assume TianQin and LISA begin observation at the same time, they share the same starting frequency. 
The characteristic strains $2f|\Tilde{h}(f)|$ from TianQin and LISA present different features, which come from the difference in their respective transfer functions. The noise $\sqrt{f S_n(f)}$ of TianQin and LISA are represented by the solid lines. Both are dimensionless so that one can make reasonable comparisons.
It is clear that response signals and noise in LISA are generally larger than those in TianQin. There are no significant differences in the total SNR, because, although the response signals of LISA in A, E, and T channels are generally higher than TianQin, the relative noise PSD is also higher as the different technical requirements.
Then, in the bottom plot of Fig.~\ref{fig:mock_data}, the GW190521 system characteristic strains show up at a lower frequency area, because the chirp mass of GW190521 is heavier and the start frequency $f_0$ is proportional to $M_c^{-5/8}$.

With the \ac{gw} strains and the PSD, the SNRs for two events are 16.1 and 14.1 (GW150914), 13.0 and 14.0 (GW190521) for LISA and TianQin. Based on the SNRs, one might conclude that LISA could provide a tighter constraint on the parameters for GW150914 while offering a weaker constraint on GW190521. 

In our work, the two \ac{gw} signals are assumed to merge within two years, which aligns with the observation time for both TianQin and LISA. However, in the 3+3 observation schemes utilized in this work, TianQin will not operate in the last three months, while LISA could observe consistently. Consequently, the high-frequency components of signals, which contribute more to the \ac{snr}, cannot be detected by TianQin. Despite this unfavorable scenario, the total SNRs obtained from TianQin are still comparable to those from LISA. 
\subsection{Parameter priors}\label{title:mock data prior}

To obtain the posterior of the parameters, one needs to specify the prior probability of all the parameters and we adopt uninformative priors therefore the prior distributions of all parameters are flat.
All the priors for the parameters are listed in Table.~\ref{tab:priors}.
In the table, the priors for $\{M_c, \eta, D_L, t_c\}$ are different for the two injected sources, the upper line is for GW150914 and the lower line is for GW190521.


\begin{table}[tbp]
    \centering
    \caption{Priors of the \ac{gw} events' parameters.}
    \label{tab:priors}
    \begin{tabular}{cc|ccccccccc}
        \hline
        \textbf{Intrinsic} & \textbf{Priors} & \textbf{Extrinsic} & \textbf{Priors} \\
        \hline
        \multirow{2}{*}{$M_c$} & $[20, 40] M_\odot$  & \multirow{2}{*}{$D_L$} & $[50, 150]$ Mpc \\
        & $[45,85] M_\odot$ & & $[400, 1400]$ Mpc \\
        \multirow{2}{*}{$\eta$} & $[0.2,0.25]$ & \multirow{2}{*}{$\Delta t_c$} & $[\rm -1000 s,+1000s]$  \\
        & $[0.2,0.25]$ & & $[\rm -1000s,+1000s]$  \\
        \hline
        $\chi_a$ & $[-1, 1]$ & $\lambda$ & $[0, 2\pi]$ \\
        $\chi_l$ & $[-1, 1]$ & $\beta$ & $[-\pi/2, \pi/2]$ \\
        $\iota$ & $[0, \pi]$ & $\psi$ & $[0,\pi]$ \\
        & & $\phi_c$ & $[0, 2\pi]$ \\
        \hline
    \end{tabular}
\end{table}

One should note that the spin parameters $(\chi_a,\chi_l)$ are the reparametrization of the spin $\chi_1$ and $\chi_2$. Because in our practice, the strong correlation between $(\chi_1, \chi_2)$ was found and it caused a decrease in sample efficiency and led to a long computational time. Thus, we utilize the parametrization parameters $(\chi_a,\chi_l)$ to help improve efficiency and the physical boundary should be considered that component spin $|\chi_i|\leq1, (i=1,2)$. 
Besides, the priors of the chirp mass $M_c$, symmetric mass ratio $\eta$, luminosity distance $D_L$, and coalescence time $t_c$ were restricted to around the injected parameters' value, since we assume the detection pipeline can successfully identify the signal, and under this assumption, we are able to start the sampling process around the injected value to boost convergence efficiency.

\subsection{Parameter estimation results}
\label{title:PE results}

This work considered an ideal scenario where only one \ac{bbh} \ac{gw} event was injected into the data. In this way, we have considered several scenarios, i.e., the two \ac{gw} events were measured by TianQin only, LISA only, and TianQin with LISA (joint), respectively. Additionally, we use the full $11$ parameters and the $9$ waveform, which $(\chi_1,\chi_2=0)$ to obtain and compare the PE results.

\subsubsection{ GW190521 PE result} \label{subtitle:individual results 521}

\begin{figure*}[htbp]
    \includegraphics[width=\linewidth]{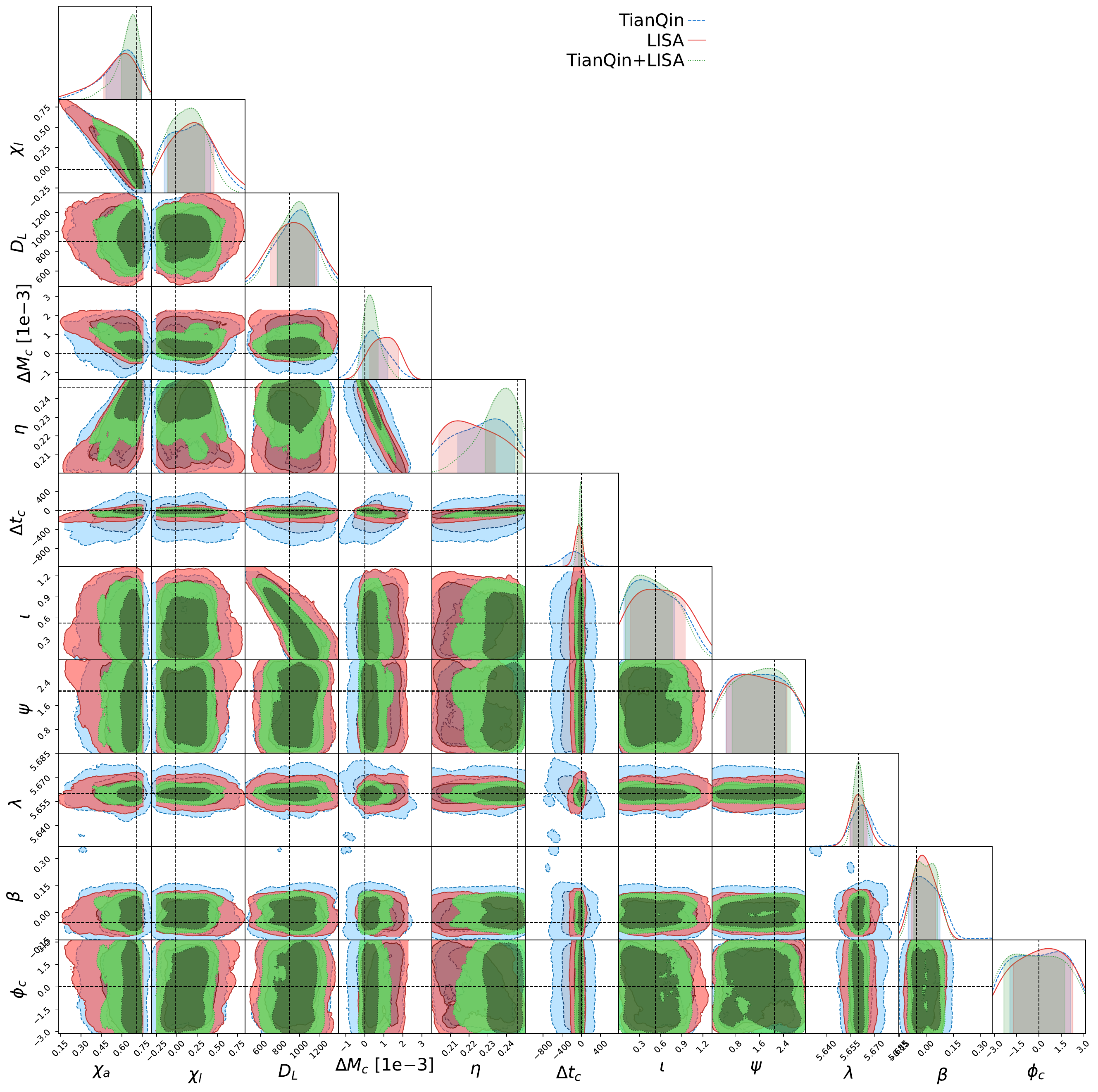}
    \caption{The combination of PE results of TianQin, LISA, and joint observation (GW190521). The black dashed lines indicate the \ac{gw} waveform parameters' true values and the $1-2 \sigma$ contour plots represent the inferred PE results. The results were obtained with our three different samples, TianQin in blue, LISA in red, and TianQin+LISA in green. The chirp mass $M_c$ and coalescence time $t_c$ are changed into the deviation value $\Delta M_c, \Delta t_c$ from their true figure.
    }
    \label{fig:multi_GW190521}
\end{figure*}

\begin{table*}[htbp]
    \centering
    \caption{1 $\sigma$ confidence region of parameters for the GW190521-like case.  
    }
    \label{tab:GW190521 PE}
    \begin{tabular}{ccccc}
        \hline
        Parameters & TianQin & LISA & Joint \\
        \hline
        $M_c (M_\odot)$ & $65.10_{-6.077\times 10^{-4}}^{+8.087\times 10^{-4}}$ &$65.10_{-4.53\times 10^{-4}}^{+7.43\times 10^{-4}}$&$65.10_{-3.48\times 10^{-4}}^{+3.47\times 10^{-4}}$ \\
        $\eta$ & $0.228_{-0.017}^{+0.014}$ &$0.224_{-0.016}^{+0.017}$& $0.238_{-0.0010}^{+0.0085}$ \\
        $\chi_a$ & $0.60_{-0.16}^{+0.098}$ & $0.61_{-0.17}^{+0.090}$ & $0.67_{-0.089}^{+0.051}$ \\
        $\chi_l$ & $0.15_{-0.30}^{+0.26}$ & $0.17_{-0.26}^{+0.28}$ & $0.082_{-0.22}^{+0.24}$  \\
        \hline
        
        $\iota(\rm rad)$ & $0.48_{-0.34}^{+0.40}$& $0.55_{-0.38}^{+0.38}$& $0.44_{-0.31}^{+0.38}$ \\
        $D_L(\rm Mpc)$ & $972.18_{-222.44}^{+178.50}$ & $962.05_{-221.23}^{+205.49}$ & $972.18_{-202.52}^{+143.45}$  \\
        $t_c-6.31\times 10^7 (s)$ & $-154_{-210}^{+197}$ &$-59.9_{-69.7}^{+53.5}$& $-22.1_{-22.8}^{+51.2}$ \\
        $\lambda (\rm rad)$ & $5.66_{-0.0064}^{+0.0061}$ & $5.66_{-0.0049}^{+0.0049}$ & $5.66_{-0.0030}^{+0.0029}$  \\
        $\beta(\rm rad)$ &$-0.0093_{-0.074}^{+0.090}$ & $-0.014_{-0.058}^{+0.067}$& $-0.0046_{-0.060}^{+0.062}$ \\
        $\psi(\rm rad)$ & $1.51_{-1.015}^{+1.089}$& $1.55_{-1.075}^{+1.092}$& $1.67_{-1.093}^{+0.97}$ \\
        $\phi_c(\rm rad)$ &$0.0067_{-2.20}^{+2.057}$ & $0.16_{-2.17}^{+2.033}$& $-0.18_{-2.15}^{+2.23}$ \\
        \hline
    \end{tabular}
\end{table*}

The constraint results for the parameters of GW190521-like event are summarized in Table.~\ref{tab:GW190521 PE} and the corresponding posterior distributions are shown in the corner plots of Fig.~\ref{fig:multi_GW190521}.
From Table.~\ref{tab:GW190521 PE}, one can find that the constraint variances of the intrinsic parameters are similar with only TianQin and only LISA, while it becomes strict for $M_c$ and $\eta$ under the joint of TianQin and LISA. For the mass parameters $(M_c,\eta)$, the PE precision could be $\approx 10^{-5}$ and $\approx 0.07$, and the joint observation will help decrease the precision to half of the TianQin and LISA. The TianQin and LISA didn't limit the spin parameters $(\chi_a,\chi_l)$ very well because the spin effect appears in the 1.5 and 2 order PN of \ac{gw} waveform, rather than the leading order.

For parameters $(D_L,\iota)$, the joint observation doesn't improve the PE results significantly. 
Both TianQin and LISA suffers from the similar degeneracy, and the combination does not help to break the degeneracy. 
For the  sky location $(\lambda,\beta)$, the joint observation shows the advantage in giving a narrower and more precise area of the \ac{gw} sources around the injected value is $(5.66,-0.055)$. For the polarization angle $\psi$ and coalescence phase $\phi_c$, the posterior distribution is generally flat. 
It can be explained by the fact that they appear only in the phase term of the response signal. 
In the 22-mode signals, polarization angle $\psi$ and coalescence phase $\phi_c$ only affect the phase, and higher harmonics waveforms will help constrain the two parameters~\cite{Toubiana:2020cqv}.

For most of the parameters of the sources in Table.\ref{tab:GW190521 PE}, TianQin and LISA have demonstrated similar constraining abilities, which show compatible $1\sigma$ widths. This is expected since the sources' SNRs are also comparable. 

While the PE results of $t_c$ show a significant divergence. The contribution of PE precision of joint observation mostly came from LISA, while TianQin is almost one order worse. This is mainly due to the fact that the time period we set. For the observation windows $2\times(3 \rm months)$ during the whole $2$ years, TianQin missed the last three month data, when the sBBHs are actually closing to merge and the frequency and waveform phase change dramatically. 
In the first order of the $t(f)$ in frequency domain waveform~\cite{Krolak:1995md, Buonanno:2009zt}, $t(f) = t_c - \frac{5}{256}M_c^{-5/3}(\pi f)^{-8/3}$  and consequently $f^{8/3} \propto (t_c - t)^{-1}$. When time is close to $t_c$, the waveform frequency and the corresponding phase will evolve dramatically compared with the time period long before the merge. Under this situation, a tiny shift in the waveform will create a great difference in likelihood value, making LISA and joint observation more sensitive to $t_c$.  
Other than the most pessimistic case we assumed for TianQin, we also tested the optimistic case for TianQin that the last three-month signal can be observed. We find the PE precision of $t_c$ for TianQin could approach about $3.9$s, which is also consistent with the PE results in \citet{liu_science_2020}. More details are presented in Appendix.\ref{appendix:A}.



The posterior distribution of whole parameters is shown in Fig.~\ref{fig:multi_GW190521}. As predicted in Sec.~\ref{title:sBBH system}, the correlation between parameters can be forecasted. From the left top of the corner plot, $\chi_a,\chi_l$ shows a strong negative correlation. In the phase of the \texttt{IMRPhenomD} waveform, the spin parameters $\chi_a$ and $\chi_l$ appear in the form of the product, while $\chi_a$ and $\chi_l$ are only the linear combinations of individual spin magnitude, could not break the degeneracy between $\chi_1$ and $\chi_2$ as we tested.
Meanwhile, the marginal correlation between the spin $(\chi_a,\chi_l)$, symmetric mass ratio $(\eta)$, and chirp mass $(M_c)$ could also be seen in the corner plot, which is consisting of the results in \cite{Baird:2012cu} that there is a correlation between mass and spin. 
For the parameter group $(M_c,\eta)$, the contour plot also shows a strong correlation, which is in good agreement with \cite{Toubiana:2020cqv}. 

We could also see a significant correlation between luminosity distance $(D_L)$ and inclination angle $(\iota)$. These two parameters only appear in the amplitude item. 
The criteria for matching waveform templates with data is the value of the product of the likelihood and prior.
The $D_L$ and $\cos{\iota}$ will only affect the amplitude of the waveform, the matching could only constrain the combination of them as a scale factor. 
The degeneracy can be observed in the TianQin and LISA PE results, and it still can not be broken under the TianQin+LISA observation. 
However, the TianQin+LISA can give a tighter constraint on $D_L$, as the 90$\%$ confidence interval (CI) of TianQin+LISA is around $346$Mpc and TianQin, LISA are about $400$Mpc and $446$Mpc respectively. Under the present methodology, it is hard to give a precise limitation on each parameter.  


\subsubsection{ GW150914 PE result} \label{subtitle:individual results 914}

\begin{figure*}[htbp]
    \includegraphics[width=\linewidth]{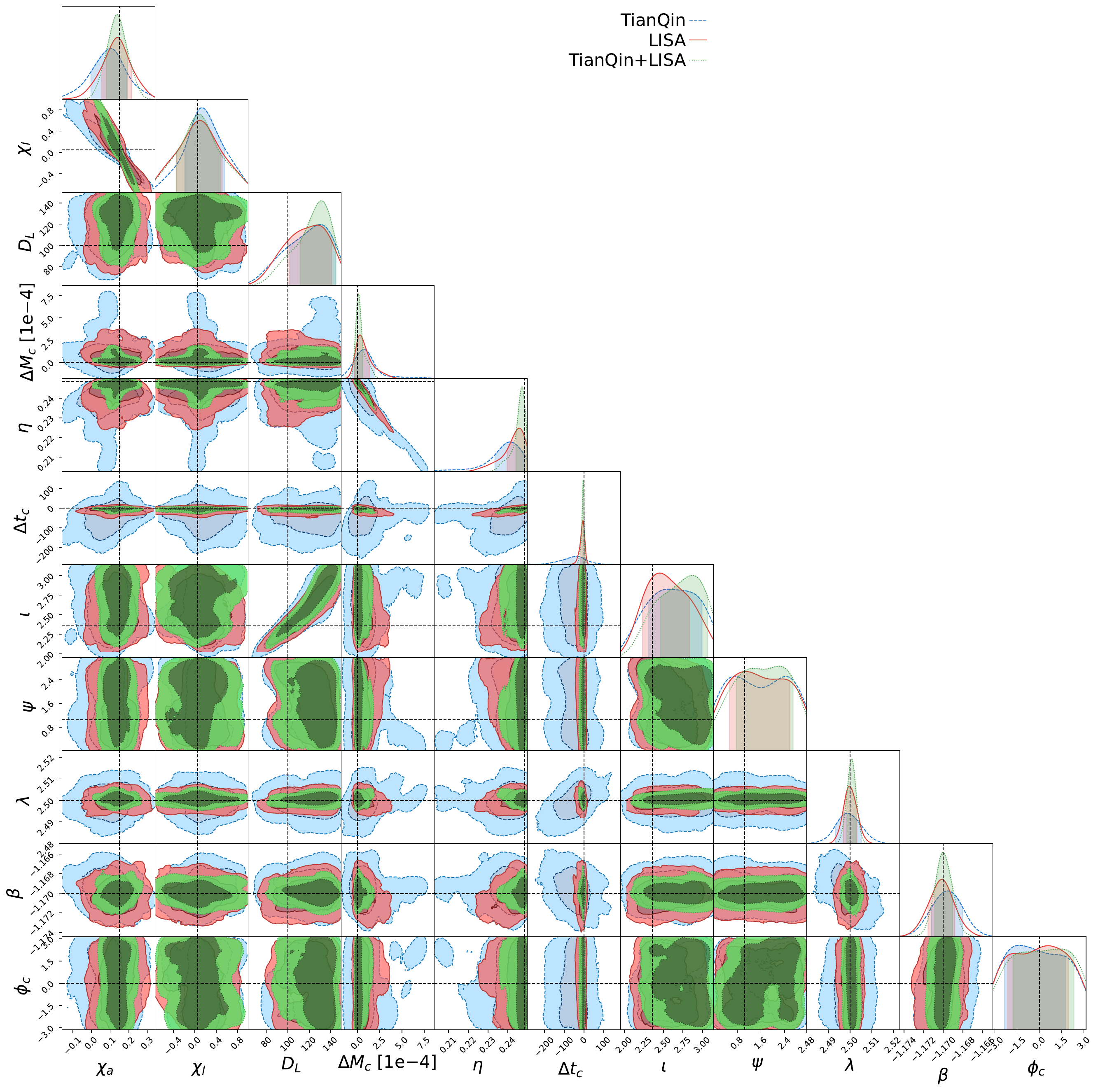}
    \caption{ The combination of PE results of TianQin, LISA, and joint observation (GW150914). The black line shows the waveform parameters' true values. The red contour line represents the posterior distribution derived by TianQin, the blue contour line indicated the posterior distribution from LISA, and the magenta contour line is the posterior distribution of joint observation. The vertical dashed lines represent the 90 percent confidence interval.
    }
    \label{fig:multi_GW150914}
\end{figure*}

The constrained results and corresponding contour plots for the parameters of GW150914-like event are summarized in Table.~\ref{tab:GW150914 PE} and Fig.~\ref{fig:multi_GW150914}, respectively.
Similar to Sec.~\ref{subtitle:individual results 521}, we will discuss the results.
\begin{table*}[htbp]
    \centering
    \caption{Constraint results of the parameters for the GW150914-like case. 
    }
    \label{tab:GW150914 PE}
    \begin{tabular}{ccccc}
        \hline
        & TianQin & LISA & Joint \\
        \hline
        $M_c$ & $28.72_{-9.77\times 10^{-5}}^{+7.81\times 10^{-5}}$ & $28.72_{-4.99\times 10^{-5}}^{+1.12\times 10^{-4}}$ & $28.72_{-2.33\times 10^{-5}}^{+6.76\times 10^{-5}}$ \\
        $\eta$ & $0.24_{-8.35\times 10^{-3}}^{+6.28\times 10^{-3}}$ & $0.244_{-7.74\times 10^{-3}}^{+3.36\times 10^{-3}}$& $0.247_{-4.059\times 10^{-3}}^{+1.51\times 10^{-3}}$ \\
        $\chi_a$ & $0.095_{-0.091}^{+0.073}$ & $0.14_{-0.077}^{+0.067}$ & $0.14_{-0.053}^{+0.054}$ \\
        $\chi_l$ & $0.16_{-0.29}^{+0.39}$ & $0.091_{-0.41}^{+0.39}$ & $0.079_{-0.41}^{+0.40}$  \\
        \hline
        $\iota$ & $2.62_{-0.32}^{+0.36}$& $2.54_{-0.25}^{+0.32}$& $2.75_{-0.35}^{+0.26}$ \\
        $D_L$ & $121.43_{-24.52}^{+17.86}$ & $118.78_{-21.10}^{+18.13}$ & $127.46_{-22.56}^{+11.23}$  \\
        $t_c - 6.31\times 10^7(s)$ & $-46.2_{-65.9}^{+42.6}$ &$-6.93_{-12.3}^{+7.47}$& $-4.59_{-8}^{+3.37}$ \\
        $\lambda$ & $2.50_{-4.92\times 10^{-3}}^{+5.78\times 10^{-3}}$ & $2.50_{-2.75\times 10^{-3}}^{+3.046\times 10^{-3}}$ & 
        $2.50_{-2.11\times 10^{-3}}^{+1.88\times 10^{-3}}$  \\
        $\beta$ &$-1.17_{-1.5046\times 10^{-3}}^{+1.34\times 10^{-3}}$ & $-1.17_{-1.30\times 10^{-3}}^{+1.16\times 10^{-3}}$& $-1.17_{-7.77\times 10^{-4}}^{+8.91\times 10^{-4}}$ \\
        $\psi$ & $1.63_{-1.18}^{+1.083}$& $1.51_{-0.96}^{+1.15}$& $1.63_{-0.93}^{+0.97}$ \\
        $\phi_c$ &$-0.062_{-2.14}^{+2.20}$ & $0.037_{-2.30}^{+2.051}$& $0.26_{-2.32}^{+2.13}$ \\
        \hline
    \end{tabular}
\end{table*}
What was noteworthy is that the PE of the spin parameters shows a divergence in three cases in Table.\ref{tab:GW150914 PE}. 
The joint observation nearly didn't improve the constrain precision, especially the $\chi_l$, which is the $(\chi_1-\chi_2)/2$. 
Comparing the initial injected values of parameters for the two \ac{gw} events, the injected value of spin parameters is much closer to zero in the case of the GW150914-like event than that of the GW190521-like event. This may suggest that the magnitude of spins for the system will affect the results of constrained variance and for lower spin magnitude \ac{gw} sources, TianQin and LISA are less capable to give a precise PE and neither for joint observation. And for mass parameters $(M_c,\eta)$, the joint observation will marginally improve the PE precision and there is also a slight improvement compared with GW190521 PE results which can be explained by the higher SNR. 
The precise estimation results of chirp mass also benefit the ability to constrain the coalescence time. Because the initial frequency primarily depends on the binary's chirp mass and coalescence time, the shift on $t_c$ will affect the frequency regions of the waveforms and naturally change the inner product value with observation data. Consequently, the TianQin and LISA are sensitive to the coalescence time. 
For sky location parameters $(\lambda,\beta)$, there is no significant improvement in joint observation. It may be the reason that the short observation time $(2\rm yrs)$ has less ability to give a better location and the results in \cite{Toubiana:2020cqv} demonstrate the longer time $(\rm 5yrs,10yrs)$ period will increase the PE precision. 
More PE details are shown in Fig.~\ref{fig:multi_GW150914}. For $D_L$ and $\iota$, the joint observation will lead to a more off-center feature while a strong correlation still exists between parameters, and the 90CI is narrower than TianQin and LISA. Meanwhile, for most of the parameters, TianQin+LISA could help give a more narrow distribution and center on the true value of the parameter.

\subsubsection{ Spin mismodeling effect } \label{subtitle: spin mismodeling}

\begin{figure*}[htbp]
    \centering
    \includegraphics[width=1\linewidth]{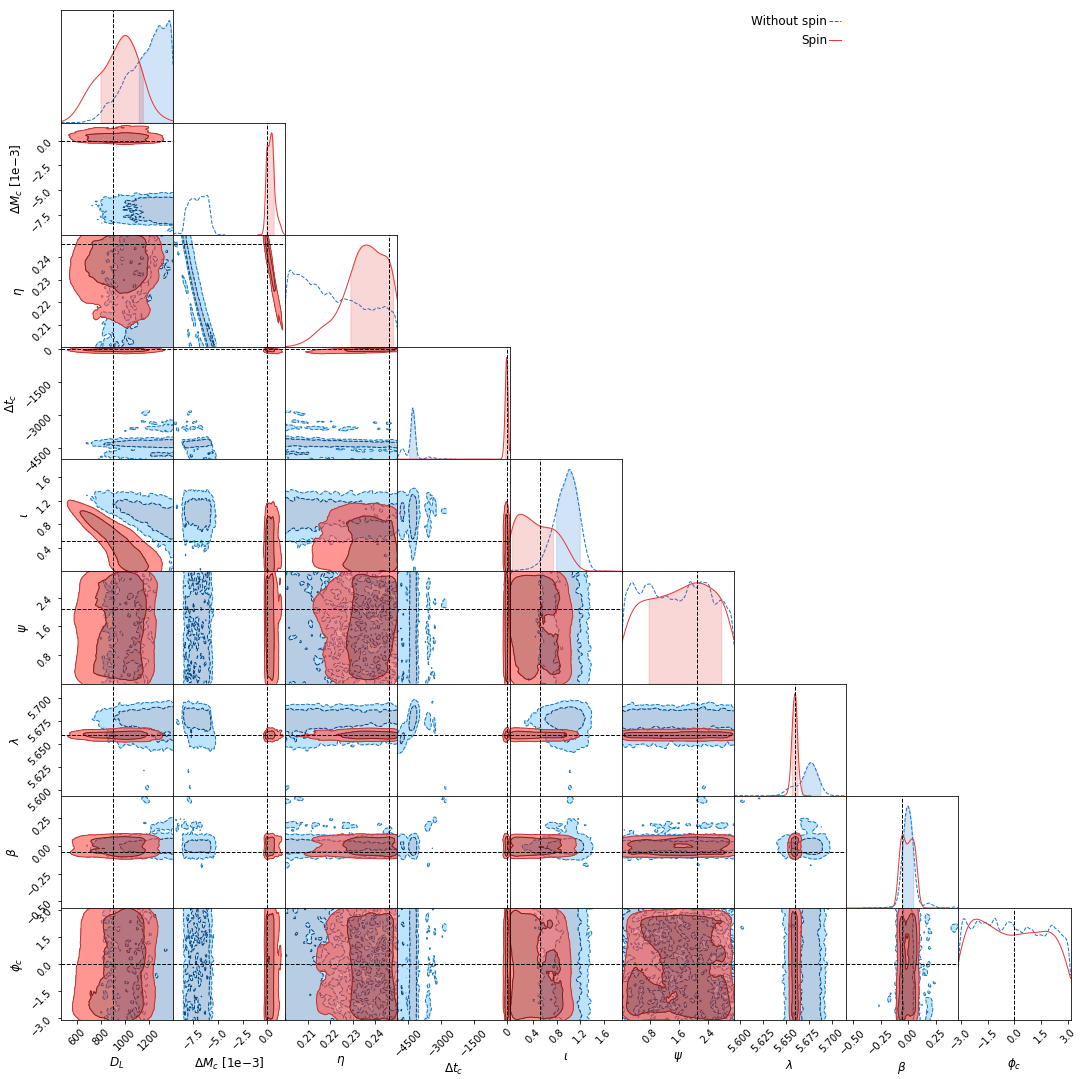}
    \caption{ The comparison between GW190521 PE results with considering the spin (spin) in red and PE results with assuming all spin to be $0$ (Without spin) in blue. The black dashed lines indicate the true value and the $1-2 \sigma$ contour. 
    }
    \label{fig:joint_GW190521_9_spin}
\end{figure*}

\begin{figure*}[htbp]
    \centering
    \includegraphics[width=1\linewidth]{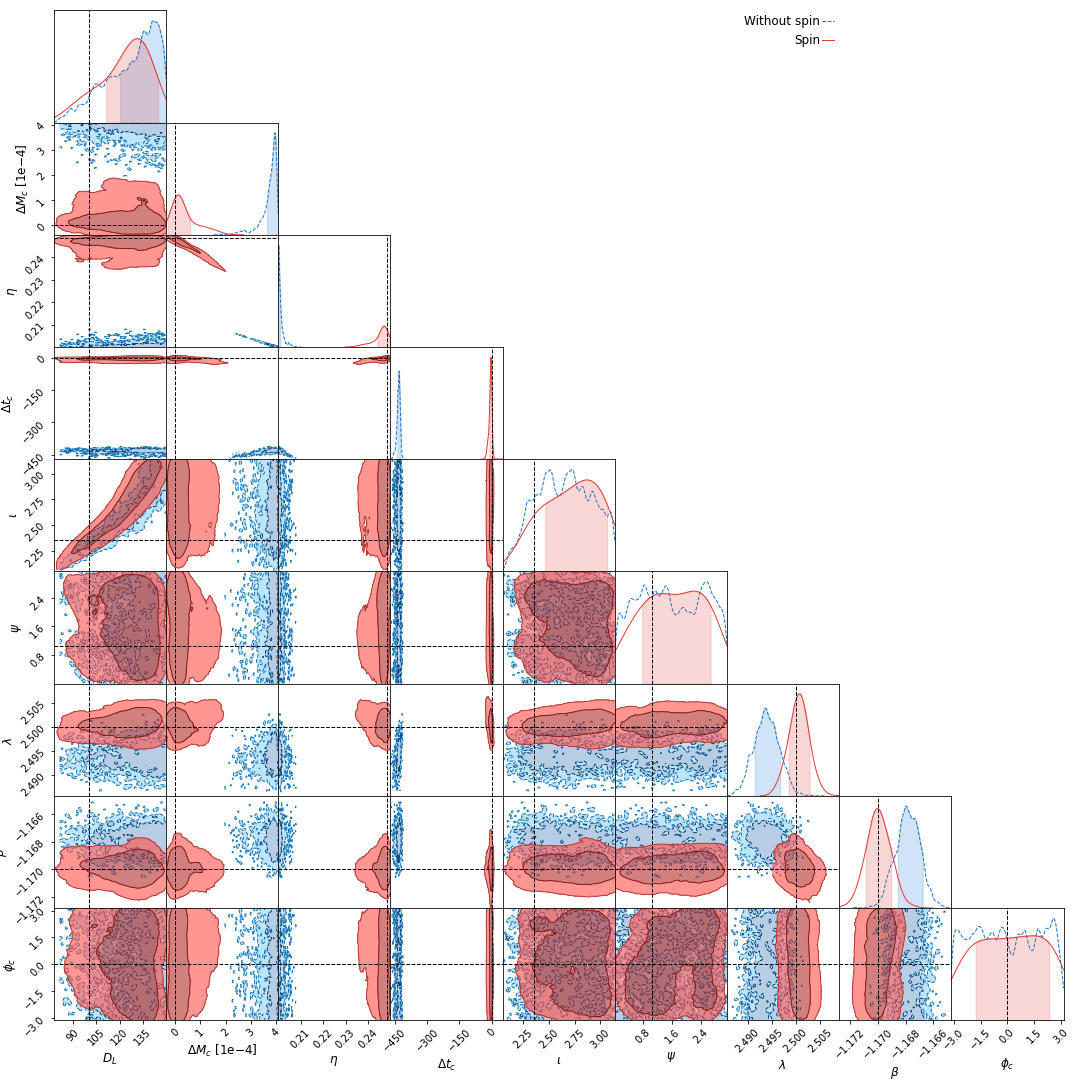}
    \caption{The comparison between GW150914 PE results with considering the spin (spin) in red and PE results with assuming all spin to be $0$ (Without spin) in blue. The black dashed lines indicate the true value and the $1-2 \sigma$ contour. 
    }
    \label{fig:joint_GW150914_9_spin}
\end{figure*}


Spin is a crucial parameter in describing the GW waveform, and the spin value can provide evidence for exploring the formation channels of sBBHs. Therefore, obtaining accurate parameter estimation results for spin is essential.

In our analysis, we investigate a common situation that can lead to biased parameter estimation results, which is spin mismodeling. Spin mismodeling means that we estimate the parameters of an \ac{bbh} source under the assumption that it has no spin, while in reality, the source has a nonzero spin.

To analyze the spin mismodeling effect on our PE results, we assume a waveform with fixed spin magnitude $(\chi_1,\chi_2)_{\rm model}=(0,0)$, while allowing the other parameters to be freely explored. Additionally, to mimic the potential differences that could arise from different detectors, we perform the joint observation (TianQin+LISA) to estimate the effect of spin mismodeling based on the same data set used in Sec.~\ref{subtitle:individual results 521} and \ref{subtitle:individual results 914}.




The TianQin+LISA parameter estimation results presented in Fig.~\ref{fig:joint_GW190521_9_spin} and Fig.~\ref{fig:joint_GW150914_9_spin} reveal a significant deviation between the posterior distribution and the true value of the GW signal for GW150914 and GW190521 system. 
This discrepancy can be anticipated, as neglecting the spin parameter leads to a noticeable difference in the likelihood value between the mismodeling waveforms and the true signal. 
Specifically, the joint distribution of mass parameters $(M_c,\eta)$ shifts far away from the $3\sigma$ region of the true parameters' value, when the spin is neglected. It indicates that mass and spin are degenerate and the absence of spin will cause a significant deviation in mass. Since mass plays a crucial role in sBBH formation, large deviations caused by mismodeling can affect subsequent analyses related to the formation mechanism~\cite{baibhav_mass_2020}, event rates~\cite{KAGRA:2021duu}, and the environment \ac{bbh} grow \cite{Zevin:2017evb,baibhav_mass_2020,Taylor:2018iat,Wysocki:2018mpo,Roulet:2018jbe}.

Moreover, \ac{bbh} systems are expected to be dark in the EM channel \cite{McKernan:2019hqs,Graham:2020gwr}, meaning that the cosmological information \ac{bbh} provided by sBBHs is limited to the GW source's sky location and luminosity distance. And the \ac{gw} sources without EM counterparts are known as “dark siren” \cite{Schutz:1986gp,DES:2019ccw}. The absence of spin also leads to an approximate $2\sigma$ shift in coalescence time and sky location, which have a significant impact on determining the Hubble constant, follow-up EM observation, and ground-based GW examinations for merger moment.  While for $D_L$ and $\iota$, there are approximate $1\sigma$ deviations from the PE results with spin, possibly due to weaker constraints on these parameters.
These results underscore the significant impact that mismodeling waveforms can have on the constraint results for the physical parameters of detected \ac{gw} events.

\section{Conclusion}\label{title:conclusion}

The \ac{bbh} inspiral systems have emerged as a significant \ac{gw} source for spaceborne GW detectors.
This work first aims to verify the full-frequency response function for LISA and TianQin and use the Bayesian inference method to extract physical information from the simulated \ac{gw} data. As the ground-based \ac{gw} detectors have published 81 \ac{bbh} merger events, we use two GW150914-like and GW190521-like systems as our target, and we further explore the effect of spin mismodeling. Further, we provide a reasonable interpretation of these PE results. The whole procedure could be generalized to the other \ac{gw} sources.

Our study primarily focuses on the construction of the TianQin simulated data, and obtaining the full-frequency TDI response function derived from the TianQin orbit function for the first time.  
The construction of the TianQin simulated data is based on the ``three months on, three months off'' observation windows and is generated directly in the frequency domain, while LISA simulated data is assumed to be consistent during the observation period. And for the purpose of inclusivity for potential applications to other \ac{gw} sources, not just \ac{bbh}, we adopt the \texttt{IMRPhenomD} waveform. 
For the efficiency of PE and to decrease the dimension of the parameter space, we consider a special case (aligned spin) and as the \ac{bbh} sources stay in the inspiral stage in the mHz frequency band, it is sufficient to concentrate on the dominant mode (22 mode), and higher harmonics are neglected.

The two cases studied in this work, GW150914-like and GW190521-like systems, are chosen to demonstrate the validity of our Bayesian inference method. 
And for TianQin, the absence of the last three months' data lead to a pessimistic situation as more SNR is lost and physical characteristic is missing, which means a decrease in the ability to constrain parameter. We adopt the Bayesian inference method, and the parameter information is recovered in the form of the posterior distribution shown in corner plots and the drawback in TianQin's pessimistic situation can be seen.
Firstly, the correlation between different parameters could also be seen, especially the joint distribution of spin parameters $(\chi_a,\chi_l)$, mass parameters $(M_c,\eta)$, luminosity distance and inclination angle $(D_L,\iota)$.
With the $h_{22}$ waveform, the two parameters $(D_L,\iota)$ are degenerate as they only existed in the waveform amplitude, and future research including higher harmonics may be helpful as some works have proven the importance \cite{Arun:2007hu,Porter:2008kn,Trias:2007fp,McWilliams:2009bg}. 
We also found a marginal relation between two spins $(\chi_a,\chi_l)$ and mass $(M_c,\eta)$, which could also be demonstrated in other people's work and a strong correlation between two spin parameters $\chi_a$ and $\chi_l$.
The coalescence phase $\phi_c$ and the polarization angle $\psi$ could not be limited in all three kinds of observation (TianQin, LISA, TianQin+LISA), as they are neither constrained well in TianQin and LISA. 

The distinct discrepancy in the coalescence time $t_c$ PE precision for different detectors is remarkable, for GW150914-like system, TianQin's is $\Delta t_c = [-68.43,+46.30]$ and LISA is $\Delta t_c = [-14.49,+7.75]$. The reason is that compared with LISA's consistent observation, TianQin has a ``three month on, three month off" schedule, and \ac{bbh} system in our work evolve faster during the last three months close to the merger. The observation could provide a better ability for LISA to constrain the coalescence time as it directly affects the time-frequency evolution for \ac{bbh} systems.
We noticed with the increase of the network SNR, nearly all the joint observation PE precision results were slightly better than single detectors.
Especially, for the $\chi_l$ in GW150914 data, the joint observation was slightly worse than the single. 
In that case, the GW150914-like system was assumed to have a tiny spin and the spin effect in the GW was too weak to be distinguished, especially the poor limitation on $\chi_l$.  
We move forward to investigate the effect of spin mismodeling. 
In this way, we ignored the spin parameter $(\chi_a,\chi_l)$ in the waveform model.
The PE results demonstrated a significant deviation from the true waveform parameter. 
It means that the spin effect plays a significant role for the \ac{bbh} evolution and should not be easily ignored. 
The basis of the PE method we used, together with the response functions and the inclusive \texttt{IMRPhenomD} waveform, have proved the potential that we could extend this process to many different \ac{gw} sources. 

\begin{acknowledgments}
    Xiangyu thank Shuai Liu for his great help at the start of this project and his support all along it. This work has been supported by 
    the Guangdong Major Project of Basic and Applied Basic Research (Grant No. 2019B030302001),
    the Natural Science Foundation of China (Grant No. 12173104),
    and the Natural Science Foundation of Guangdong Province of China (Grant No. 2022A1515011862). 
\end{acknowledgments}

\begin{appendix}

\section{Impact of the coalescence time} \label{appendix:A}

The 3+3 observation mode of TianQin means that if the binary coalesces in the non-observing period, then a significant part of the high SNR data will be missed.
  In this appendix, we demonstrate the impact of the choice of coalescence time $t_c$ to the  PE results for TianQin.

In our work, we fix $t_c$ to be two years for both GW150914-like and GW190521-like systems, meanwhile the total observation period for TianQin and LISA is also set to two years. We assume that LISA observes continuously for two years, while TianQin observes in a 3+3 mode. 
In our observation period setting, TianQin will start observation from begin to the third month, shut down in the next three months, and start observation again.
This means that the merger happens at the end of the non-observing three-month period. 
Since the last three month data embodies the strongest SNR, and results shown in Fig.~\ref{fig:multi_GW190521} and Fig.~\ref{fig:multi_GW150914} correspond to the pessimistic scenario. 
This setting is quite different from the optimistic setting in other works like \citet{liu_science_2020}, where $t_c$ is set to the beginning of the non-observing three-month period.
  Comparisons between our MCMC results to Fisher analysis should be treated carefully to include this difference.

\begin{figure}[htbp]
    \centering
    \includegraphics[width=0.9\linewidth]{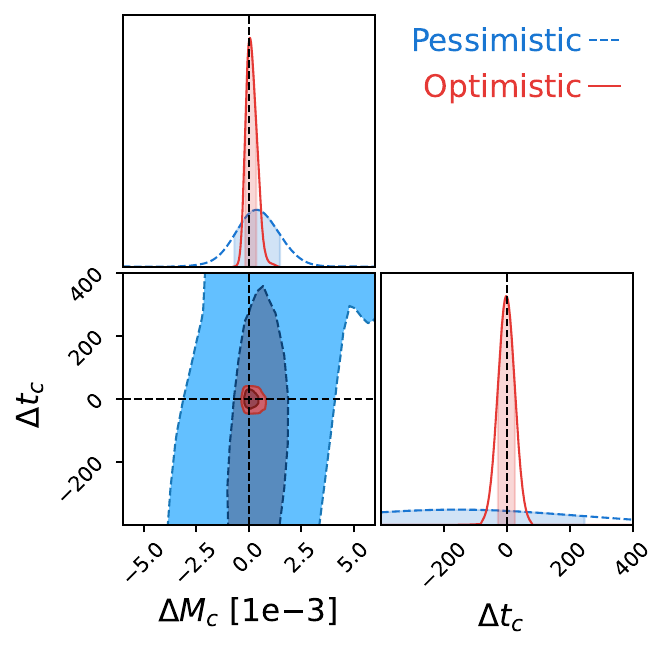}
    \caption{ The comparison between the optimistic and pessimistic PE results of TianQin. The black dashed lines, dark blue area and light blue area indicate the true value and the $1-2 \sigma$ contour, respectively. 
    }
    \label{fig:pess_opti}
\end{figure}

In order to make a more direct comparison with \citet{liu_science_2020}, for the GW190521-like event, we set the $t_c$ at the beginning of the last non-observing three-month, but keep every other parameter fixed.
We then perform parameter estimation on the new data and present the comparison of results from different $t_c$ in Fig.~\ref{fig:pess_opti}.
The coalescence time $t_c$ can be constrained with a precision of around $3.9$s, which is comparable with \citet{liu_science_2020} and is better than LISA's.
For this optimistic scenario, the last three month data will be detected by TianQin, and the fast frequency evolution will also dramatically improve the precision on the mass parameter.
We notice that by shifting the $t_c$, the precision on the chirp mass $\mathcal{M}_c$ improves from $1.42\times10^{-3}M_\odot$ to $4.86\times10^{-4}M_\odot$.

\end{appendix}

\bibliography{refpaper} 

\end{document}